\documentclass[]{interact}

\usepackage{graphicx}       
\usepackage{amsmath, amssymb}
\usepackage{color,xcolor}
\usepackage{tikz}
\usetikzlibrary{arrows}
\usepackage{times}
\usepackage{url}
\usepackage[utf8]{inputenc}
\usepackage{fontawesome5}
\usepackage{adjustbox}
\usepackage[small]{caption}
\usepackage{subcaption}

\usepackage[natbibapa,nodoi]{apacite}
\usepackage{epstopdf}
\usepackage{comment}
\usepackage{lineno}

\usepackage{amsthm}
\theoremstyle{plain}

\theoremstyle{definition}

\theoremstyle{remark}

\usepackage[hidelinks]{hyperref}

\begin{document}
 
\title{On using AI for EEG-based BCI applications: problems, current challenges and future trends 
}

\author{
\name{Thomas Barbera\textsuperscript{a}\thanks{CONTACT T. Barbera. Email: thomas.barbera@unimib.it}, 
Jacopo Burger\textsuperscript{b},
Alessandro D'Amelio\textsuperscript{b},
Simone Zini\textsuperscript{a},\\
Simone Bianco\textsuperscript{a},
Raffaella Lanzarotti\textsuperscript{b},
Paolo Napoletano\textsuperscript{a},\\
Giuseppe Boccignone\textsuperscript{b},
Jose Luis Contreras-Vidal\textsuperscript{c}}
\affil{
\textsuperscript{a}University of Milano-Bicocca, Italy; 
\textsuperscript{b}University of Milan, Italy;
\textsuperscript{c}University of Houston, TX, USA}
}

\maketitle

\begin{abstract}

Imagine unlocking the power of the mind to communicate, create, and even interact with the world around us. Recent breakthroughs in Artificial Intelligence (AI), especially in how machines ``see'' and ``understand'' language, are now fueling exciting progress in decoding brain signals from scalp electroencephalography (EEG). \emph{Prima facie}, this opens the door to revolutionary brain-computer interfaces (BCIs) designed for real life, moving beyond traditional uses to envision Brain-to-Speech, Brain-to-Image, and even a Brain-to-Internet of Things (BCIoT).

However, the journey is not as straightforward as it was for Computer Vision (CV) and Natural Language Processing (NLP). Applying AI to real-world EEG-based BCIs, particularly in building powerful foundational models, presents unique and intricate hurdles that could affect their reliability.

Here, we unfold a guided exploration of this dynamic and rapidly evolving research area. Rather than barely outlining a map of current endeavors and results, the goal is to provide a principled navigation of this hot and cutting-edge research landscape. We consider the basic paradigms that emerge from a causal perspective and the attendant challenges presented to AI-based models.  Looking ahead, we then discuss promising research avenues that could overcome today's technological, methodological, and ethical limitations. Our aim is to lay out a clear roadmap for creating truly practical and effective EEG-based BCI solutions that can thrive in everyday environments.
\end{abstract}


\section{Introduction}

This paper navigates the intersection of EEG-based BCI and the most recent advances in AI, to some extent, an uncharted territory. 

On the one hand, BCIs enable direct communication between the brain and external devices by interpreting neural signals. The implications of such technology are profound, offering new horizons for individuals grappling with disabilities, neurological disorders, and the everyday challenges of communication. On the other hand, the recent wave of AI advancements, particularly in Machine Learning (ML) / Deep Learning (DL) with large-scale models (e.g., generative and foundation models), marks a significant leap forward compared to early DL-based EEG analyses  (cfr. \citep{craik2019deep} for an in-depth review)  and it holds promise to potentially reshape our approach to mine such complex data and to unveil patterns within vast and nuanced datasets.

Yet, as we stand on the brink of unprecedented improvements in the interpretation of brain activity, issues are raised that are not merely a technical feat. Decades of cognitive neuroscience studies have primarily been limited to recording
the brain activity of immobile participants in a laboratory setting, most often with static stimuli. Technological advances have enabled the development of miniaturized, wireless, and wearable scalp EEG equipment that allows mobile recordings of electrophysiological brain activity, making it a promising method to study brain activity during naturalistic movement and behavior \citep{pacheco2024dance}, \citep{Theofanopoulou2024dance}. ``Mobile cognition'', by enhancing ecological validity, holds the potential to transform cognitive neuroscience by enabling the investigation of neural mechanisms and natural human behavior in dynamic, real-world environments. However, the shift towards ecologically valid approaches is very challenging in several respects (both methodological and operational, but see \citep{vigliocco2024ecological} for an in-depth discussion). In short, the research cycle involving the stages of ``Taking the laboratory into the real-world'' and ``Taking the real-world into the laboratory'' \citep{vigliocco2024ecological} requires methodological re-thinking, complex experimental designs and addressing novel ethical problems, while producing multidimensional data that are hard to analyze and interpret. This latter point, \emph{prima facie}, might ideally match the capabilities of current large-scale DL models. However, the generalizability aimed at, and achievable in principle by mobile cognition approaches might paradoxically hinder the interpretability and generalizability of results obtained via large-scale DL models.

Under such circumstances, crucial questions arise. Given the complex landscape of current EEG-based BCI tasks/paradigms/applications, can we characterize in a principled way the main challenges that current AI/ML models are likely to face? In particular,   are these models adequate to address  experimental designs and factors that might limit the performance of AI-based BCI systems when dealing with real-world, out-of-the-lab  scenarios?  Do currently employed EEG datasets support the data challenges arising in such scenarios? What aspects might undermine the interpretability and generalization capabilities of AI-based systems?

Untangling the knot of these intricate questions, demands more than a simple overview of ML approaches currently pursued in the BCI arena \citep{Edelman2024BCI}. Indeed, differently from fields where AI/ML has proved to be highly effective, e.g., natural language processing and computer vision, here the generation of the EEG data to be mined is the final step of a complex process in which several factors play a causal role: from the experimental/contextual  environment, the stimuli presented to the subject, the individual's brain activity, down to the recording mode. Figure \ref{fig:map} provides at a glance a principled view of such  a complex landscape. 

This paper unfolds as follows. By embracing a causal perspective, we first chart the landscape of EEG-based BCI paradigms according to core dimensions of stimuli and subject's engagement (Section \ref{sec:Model}).
This provides a principled map of the landscape which allows for identification of the different scenarios. Each scenario or paradigm can be a source of possible shifts in the distributions concerning its fundamental factors, the first grand challenge in this field (Section \ref{sec:context}). The second grand challenge (Section \ref{sec:data}) is represented by the data problem concerning both the data scarcity aspect, which  we formalize to allow for a quantitative comparison against LLMs, and the aspect   involving the lack of interventional/contextual data. Such principled mapping and consequential criticalities suggest future trends that are likely to develop in the effort to bridge current technological and methodological gaps, which we eventually discuss in Section \ref{sec:trends}.

\section{Charting the landscape of current approaches}
\label{sec:Model}

By and large, the application of AI techniques to BCI entails the adoption of data-driven statistical models whose accuracy depends on their complexity and the availability of sufficiently large and diverse datasets. However, when employed outside of a controlled laboratory setting, BCI applications tend to exhibit severe limitations due to poor out-of-domain data distribution shift robustness. Indeed, 
the generalization problem appears to be one of the major challenges hindering the deployment of BCIs in the real-world \citep{gao2021interface}.  As a matter of fact, the data distribution of the observed EEG traces in one dataset is likely to significantly vary in another. The driving factors of such distribution shifts can be identified by embracing a causal view on the brainwave modeling problem \citep{scholkopf2021toward,barmpas2024causal}. 

\begin{figure}[tb]
\centering
    \includegraphics[width=0.85\linewidth]{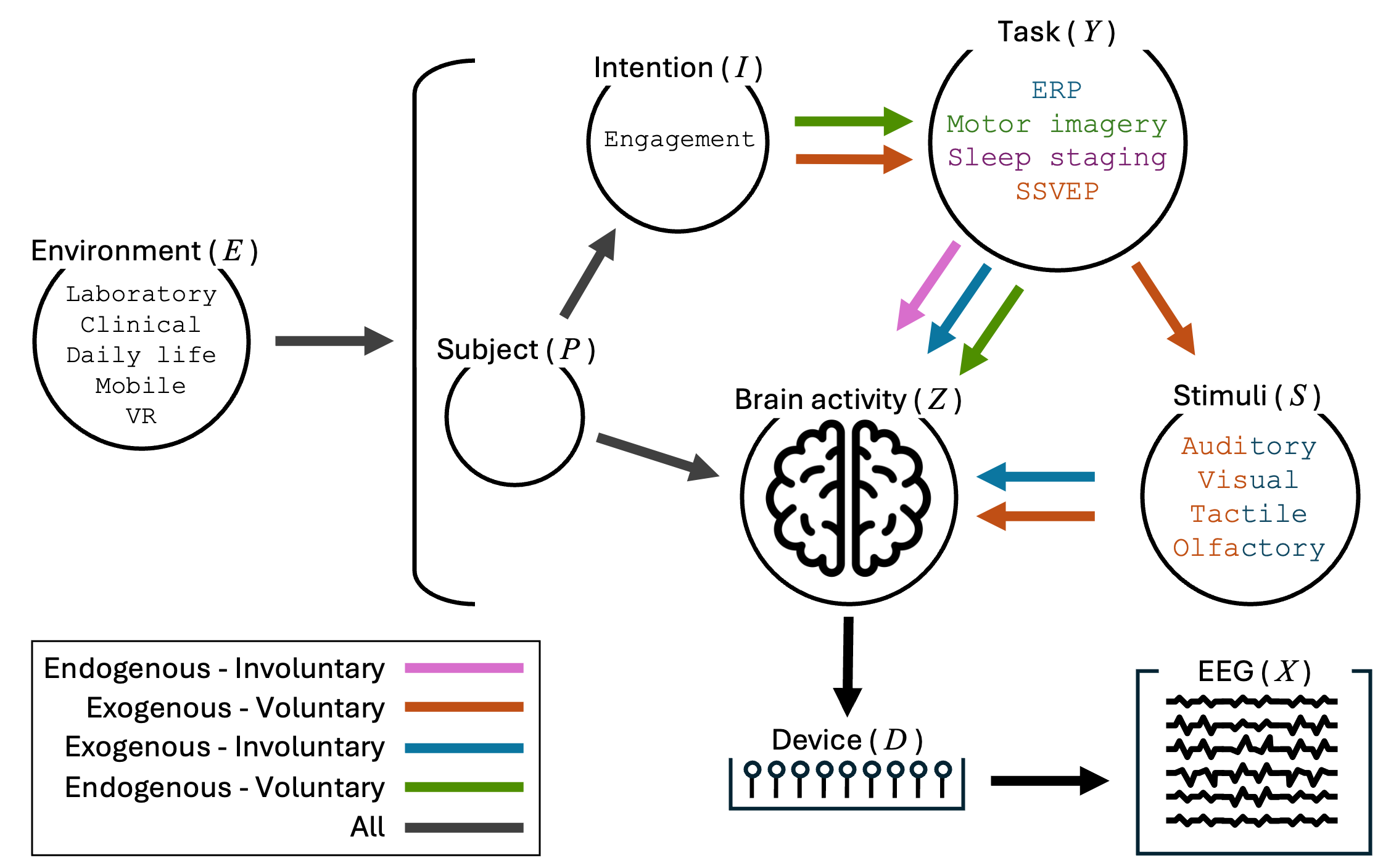}
    \caption{Causal graph describing the EEG-based BCI data generating process. Different colored arrows indicate the four scenarios specified in the bottom-left panel and set the appropriate causal relationships among the relevant variables $E,P,I,Y,Z,S,D,X$. Black arrows indicate relationships valid across all scenarios.}
    \label{fig:map}
\end{figure}

EEG-based BCIs aim at capturing patterns from a brainwave signal, $X$ (measured by an EEG device $D$), representing a noisy estimate of the underlying, unobserved brain activity, $Z$, eventually subject to a stimulus $S$. We are generally interested in recognizing the condition or decoding signal $Y$ of a subject $P$. In turn, these entities may be conditioned by other factors such as the subject's level of engagement ($I$), as well as the environmental factors $E$. The overall EEG data generating process is depicted at a glance by the causal graph in Figure \ref{fig:map}. 

Cogently, as discussed by \citep{barmpas2024causal}, the causal connections between the identified variables can vary depending on the adopted experimental setting. The possible   scenarios can be categorized by resorting to two core dimensions: 1) \emph{Exogenous vs. Endogenous}, depending on whether an external stimulus $S$ taking any form of
sensory input is present or not (also referred as Evoked vs. Spontaneous); 2) \emph{ Voluntary vs Involuntary}, based on the subject’s willingness to deliberately generate brain activity (also referred to as Active 
vs. Passive).  On this basis,  BCI paradigms can be clustered according to four distinct categories as summarized via color-codes in Figure \ref{fig:map}. Notably,  each paradigm/category yields a specific causal dependency chain among the considered factors. 

As the  model represents a causal graph, each intervention on any of the considered variables has the potential to induce significant data distribution shifts in the observed EEG traces, eventually compromising AI-based BCI generalization abilities \citep{scholkopf2021toward}.
We emphasize that these shifts are far from straightforward due to several critical factors that must be carefully assessed to ensure the reliability and effectiveness of the adopted solutions.

For example, in Involuntarily engaged and Exogenous tasks, dependencies learned with a specific set of stimuli may not hold if those stimuli are altered (Stimulus shift). In Voluntarily engaged and Endogenous tasks, it is essential to ensure that all subjects share the same intent, as variations in cognitive engagement can affect performance (Shifts on subject’s intention).  Furthermore, findings from one group of individuals cannot be assumed to generalize to others (Subject shift), nor can they be expected to remain valid when the acquisition device is changed (Acquisition shift) or the BCI system leads to changes in the EEG due to continued use (i.e., neuroplasticity). These challenges underscore a critical limitation, particularly given that most datasets are collected under specific environmental conditions, while the ultimate goal is to develop robust, universal solutions that function reliably in real-world and ecological settings, where various uncontrolled factors may compromise their validity (Shifts on environmental conditions).

\subsection{A Principled Map of EEG-based BCI Tasks}
\label{sec:tasks}

EEG-based BCI applications  
are widely designed for healthcare and well-being applications, and more recently, they have also been adopted by healthy individuals in non-medical applications \citep{varbu2022past}. In the perspective of this note, they can be charted  by following the categorization previously discussed.

\subsubsection{Endogenous and Voluntary}

This category includes clinical applications designed for patients with impairments, serving either assistive or rehabilitative purposes \citep{Edelman2024BCI}.

\textit{Assistive BCI} systems aim to compensate for lost functions in daily life, such as enabling controlling a motorized wheelchair \citep{naser2023towards}, operating an assistive robot \citep{kim2024noir}, using a neuroprosthesis or for neurorehabilitation \citep{Bhagat2020EEGBCI}. These systems rely on Motor Imagery (MI) \citep{al2021deep}, a distinct brain activity elicited when an individual imagines or attempts limb movements. 

\textit{Rehabilitative BCI} systems, also known as  neurofeedback-based BCIs, leverage neuroplasticity, which enables individuals to relearn and modulate brain activity, facilitating functional recovery. Operant conditioning is achieved by providing real-time feedback, reinforcing successful brain activity \citep{mane2020bci}.

\subsubsection{Endogenous and Involuntary}

This class of BCIs is designed to passively monitor neural signals for assessing cognitive and neurological conditions, with applications spanning healthcare, safety, and human performance evaluation.

In medical diagnostics, EEG-based solutions have proven their validity in early diagnosis and long-term monitoring of disease progression. In this framework \textit{sleep staging} analyzes EEG patterns to classify sleep phases, aiding in the diagnosis of disorders such as insomnia and sleep apnea \citep{fu2021deep}. \textit{Seizure detection and monitoring} identifies epileptic activity, enabling real-time alerts and personalized treatment adjustments \citep{ein2023eeg}. Similarly, in neuro-degenerative disease research, \textit{dementia diagnosis and monitoring} leverage EEG biomarkers to assess cognitive decline, distinguishing between conditions such as Alzheimer's and fronto-temporal dementia \citep{tanaka2024neurophysiological}. 

Beyond medical applications, EEG-based BCIs contribute to safety and cognitive performance assessment. \textit{Driver drowsiness detection} monitors fatigue and alertness, playing a vital role in accident prevention for transportation and industrial settings \citep{borghini2014measuring}. Similarly, \textit{mental workload assessment} tracks cognitive load, helping to optimize task performance and mitigate cognitive overload \citep{kingphai2024mental}.

\subsubsection{Exogenous and Voluntary}

Exogenous and voluntary BCIs leverage stimuli-driven brain activity to enable applications such as communication, decision-making, and behavioral choice.

By far, a key function of this paradigm is to facilitate assistive communication, particularly for individuals with severe motor impairments. Within this context, \textit{Event-Related Potential (ERP)}-based communication systems, such as the \textit{P300 speller} \citep{won2019p300}, identify the P300 wave---an EEG response triggered when a user detects a target stimulus within a sequence of flashing options. Similarly, \textit{Steady-State Visual Evoked Potential (SSVEP)}- and \textit{Steady-State Auditory Evoked Potential (SSAEP)}-based BCIs utilize periodic visual or auditory stimuli to elicit neural oscillations at specific frequencies and brain regions \citep{aznan2019simulating}.
Additionally, \textit{Somatosensory Evoked Potentials (SEPs)} provide another input modality for BCI interaction by utilizing tactile or vibrational stimuli to elicit neural responses. 

\subsubsection{Exogenous and Involuntary}

This category includes BCI applications designed to analyze the brain's natural response to stimuli, capturing both emotional and cognitive reactions. In \textit{emotion recognition} applications, EEG signals—sometimes combined with other physiological signals in a multi-modal approach—are processed to identify affective states \citep{torres2020eeg,sorino2024ariel}; \textit{stress monitoring} aims to detect stress levels by analyzing variations in EEG features such as spectral power and connectivity patterns \citep{badr2024review}. Very recently, the adoption of foundation models trained on EEG-signals and fine-tuned for stress detection has been proposed \citep{lloyd2024stress}. Moreover, integrating emotion recognition from EEG with large language models (LLMs) enables new applications, such as emotion-aware conversational support \citep{sorino2024ariel}.

\textit{Error-Related Potentials (ERPs)} play a role in cognitive processing, automatic error detection, and attention mechanisms \citep{xavier2022error}. They enhance human-computer interaction by enabling systems to detect and respond to user mistakes without explicit input.

Recently, new applications  have become viable through the integration of EEG signal analysis with deep learning, mainly Large Language models (LLM). Notably, this includes EEG-to-text systems, even with open-vocabulary capabilities \citep{amrani2024deep}, and EEG-to-image reconstruction \citep{ren2021reconstructing}.

\section{The Challenge of  Distribution Shifts}
\label{sec:context}

Navigating the EEG-based BCI literature through the lens of the causal reasoning, highlights how most of the approaches reviewed in Section \ref{sec:tasks}, albeit not in explicit terms, focus on the specific conditions induced by choosing a given configuration of the variables in the causal model in Figure \ref{fig:map}. By doing so, these methods may overlook essential contextual information needed for proper generalization. Notably, any change in the intervention set induces a data distribution shift, violating the i.i.d. assumption in machine learning terms \citep{scholkopf2021toward}.

Recently, several efforts have focused on developing solutions to enhance the generalization capabilities of AI-based BCIs. Although these solutions may not explicitly employ causal reasoning, they aim to address specific distribution shifts by considering certain boundary conditions. These approaches can be categorized as follows.

\subsection{Data Augmentation}

The simplest way to enrich the distribution of EEG traces in the training set is through data augmentation (DA), which increases data diversity by introducing artificially generated interventions. In the EEG-based BCI literature, in addition to traditional techniques such as adding noise, applying sliding windows, re-sampling, and recombining segmented data, DA can be implemented using more advanced methods that generate synthetic data through generative ML algorithms. These include generative adversarial networks (GANs) \citep{habashi2023generative}, generative pre-trained transformers (GPTs) \citep{bird2021synthetic}, variational autoencoders (VAEs), and diffusion models \citep{zhao2024eeg}, all of which have proven effective in generating synthetic EEG data \citep{lashgari2020data,he2021data}.

The effectiveness of DA has been validated across various applications, including clinical use cases \citep{carrle2023generation}, emotion recognition \citep{zhao2024eeg}, brain-computer interfaces (BCIs) \citep{kalaganis2020data}, steady-state visual evoked potentials (SSVEP) \citep{aznan2019simulating}, and motor imagery (MI) \citep{pei2021data}.

\subsection{Acquisition and Subject-Invariant Models}

In contrast to DA, which enhances the training set by introducing artificial distribution shifts, other approaches address the generalization challenge in BCI solutions by focusing on modeling choices. 

Specifically, to mitigate acquisition and environmental shifts \citep{awais2024lab}, several studies have explored EEG denoising techniques, with a particular focus on GAN-based methods, which have proven effective in reducing these artifacts \citep{kalita2024aneeg}.  

To address subject shift \citep{song2025domain}, normalization techniques have traditionally been used \citep{he2019transfer}. More recently, transfer learning has emerged as a promising solution \citep{wan2021review}, along with zero-calibration networks, such as adversarial inference frameworks \citep{ozdenizci2020learning} and dynamic convolution models \citep{barmpas2023improving}. In a different vein, \cite{norskov2023cslp} address noise and subject variability by extraction of latent representations accounting for content (actual neural activation) and style (subjectivity) . This is achieved via a contrastive split-latent permutation autoencoder (CSLP-AE) framework that directly optimizes for EEG conversion.

\subsection{Pre-training and Self-supervision}
\label{sec:ssl}

Similar to data augmentation, the primary objective of pre-training is to enhance the diversity and richness of the training set distribution. This approach operates under the premise that exposing a machine learning model to an extensive, heterogeneous training set allows it to absorb additional insights from multiple data distributions. This enriched learning process can improve the final model’s performance and foster greater robustness to data distribution shifts in BCI applications \citep{scholkopf2021toward}.

\subsubsection{EEG Foundation Models}
Contemporary solutions adopting pre-training on EEG data predominantly leverage self-supervised learning. In an initial stage, models are trained on a vast collection of unlabeled data, where appropriate pretext tasks (typically involving spatio-temporal masking of EEG traces) are designed to provide meaningful representations for an unknown downstream task. In a subsequent stage, the obtained model is fine-tuned for a specific task using a limited number of labeled brainwave samples \citep{gramfort2021learning}.

Inspired by the success of Transformers in Natural Language Processing, recent solutions have increasingly integrated attention mechanisms \citep{abibullaev2023deep} and large-scale self-supervised pre-training to develop EEG foundation models—often referred to as Large Brain Models (LBMs), such as BENDR \citep{kostas2021bendr}, LaBraM \citep{jianglarge}, BrainWave \citep{yuan2024brant}, Neuro-GPT \citep{cui2024neuro}, BIOT \citep{yang2024biot}, and EEGPT \citep{wang2024eegpt}. These models aim to improve generalization across diverse distribution shifts by enabling cross-data self-supervised learning. Notably, they claim to account for variations such as inter-subject differences, mismatched electrode channels, variable experiment durations, diverse task designs, low signal-to-noise ratios, and missing values.

Yet, despite being proposed as a solution to the data distribution shift problem, LBMs pre-trained with self-supervision still lack strong empirical evidence supporting these claims—primarily due to the difficulty of accessing data in sufficient quantity and quality (see Section \ref{sec:data}). Moreover, it has been postulated that LBMs may benefit from causal reasoning, particularly in designing the most appropriate pretext task for a given BCI application \citep{barmpas2024lbm}.

\section{The Data Challenge}
\label{sec:data}

\begin{figure*}[t]
\centering
    \begin{subfigure}[b]{\linewidth}
    \centering
    \includegraphics[width=0.95\linewidth]{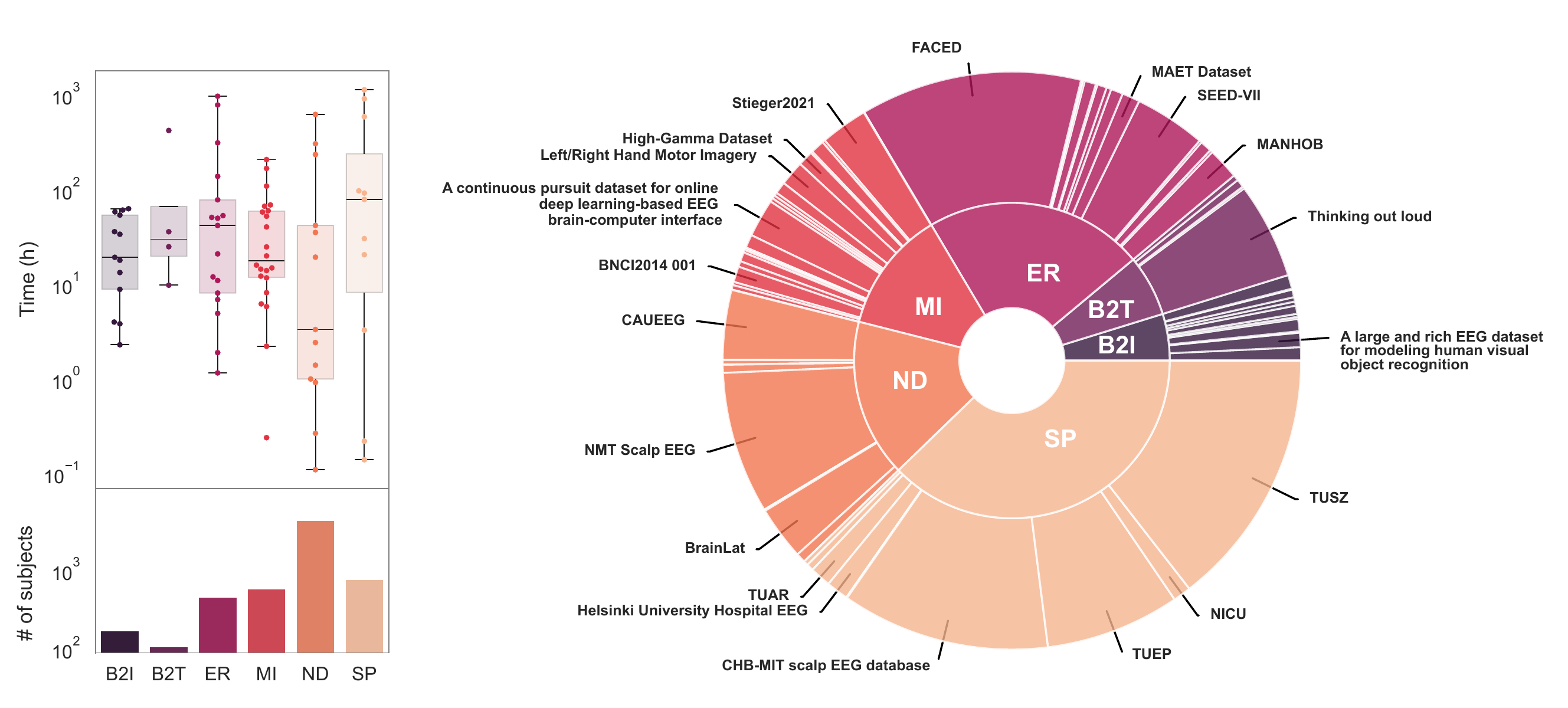}
    \caption{\label{fig:data_time}}
    \end{subfigure}

    \begin{subfigure}[b]{0.45\linewidth}
    \centering
    \includegraphics[width=\linewidth]{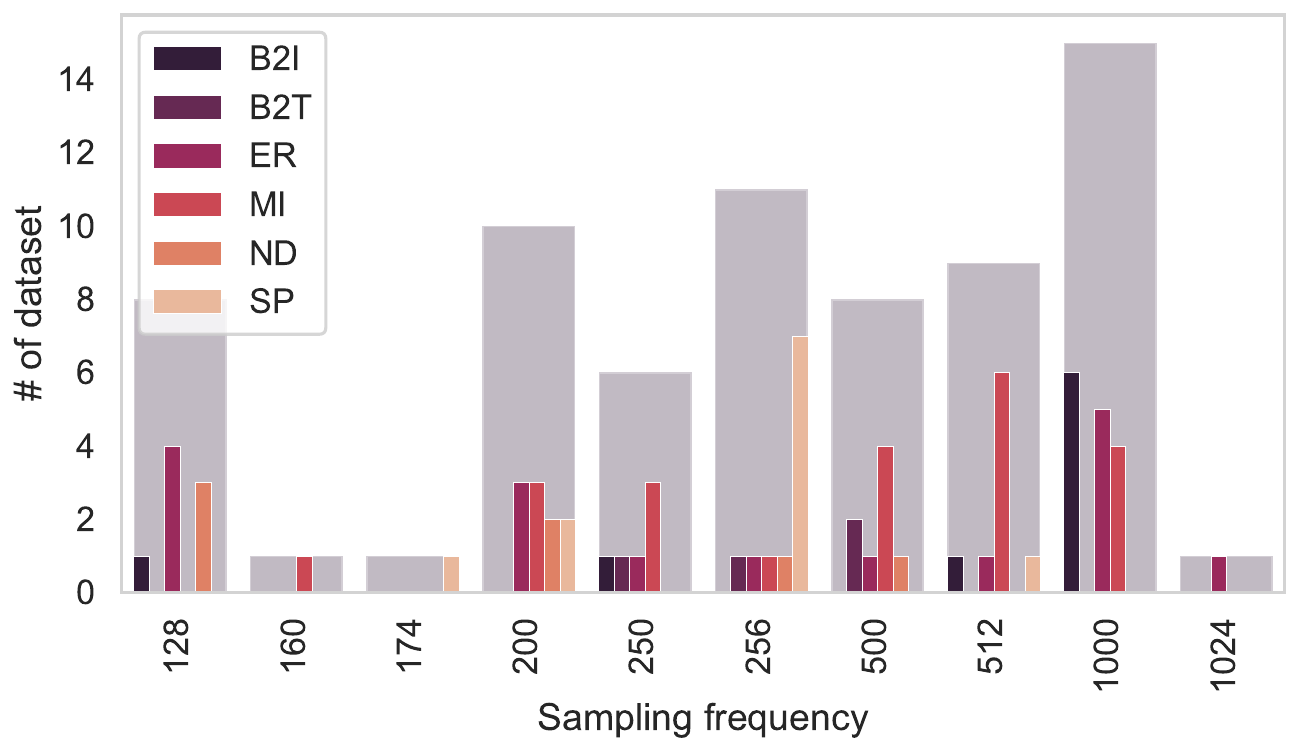}
    \caption{\label{fig:data_samplingfreq}}
    \end{subfigure} \hspace{2mm}
    \begin{subfigure}[b]{0.45\linewidth}
    \centering
    \includegraphics[width=\linewidth]{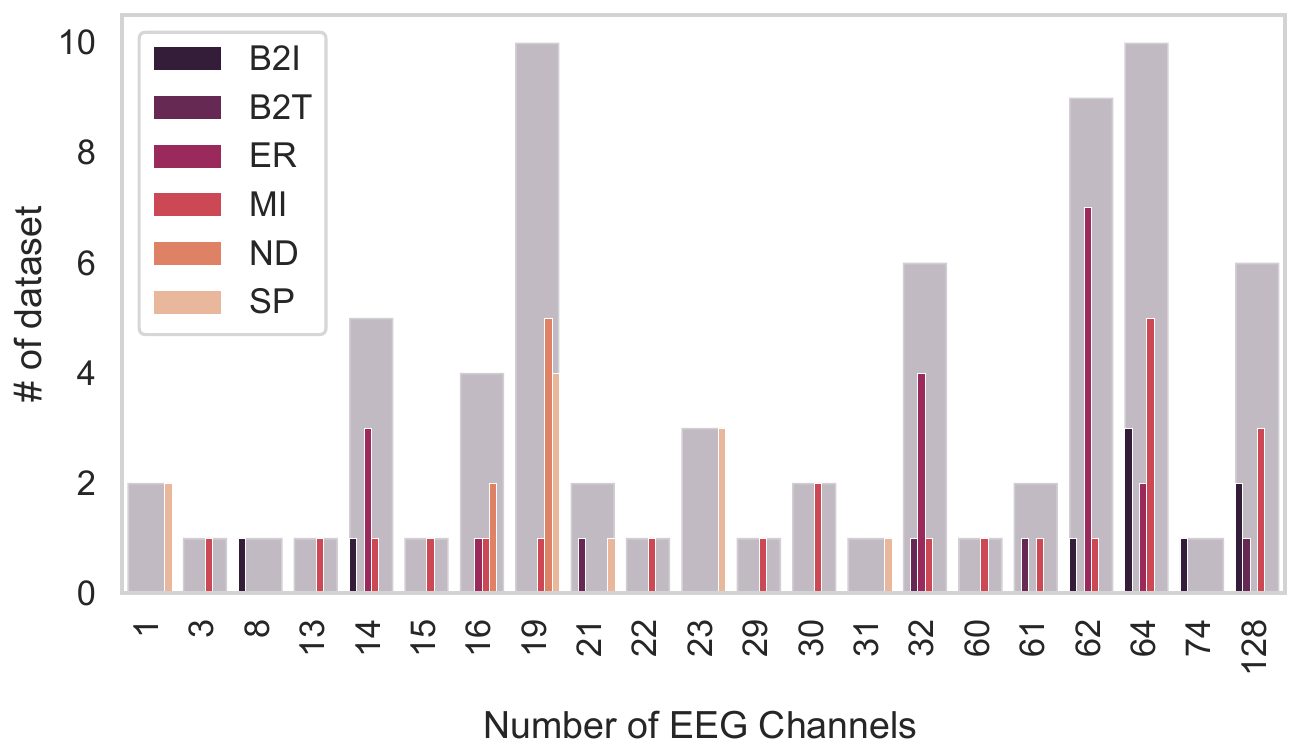}
    \caption{\label{fig:data_channels}}
    \end{subfigure}

\caption{Datasets for EEG-BCI tasks. Metadata collected for six different tasks: Brain-to-Image (B2I), Brain-to-Text (B2T), Emotion Recognition (ER), Motor Imagery (MI), Neurodegenerative Disease (ND) and Seizure Prediction (SP). (a) For each task, the total recording time (in hours) and the total number of subjects are reported: the inner circle represents the total recording hours per task, while the outer circle indicates the total recording time for each dataset per task. (b) Distribution of sampling frequencies used for collecting each dataset, categorized by task. (c) Distribution of the number of EEG channels used for collecting each dataset, categorized by task. A table with all the datasets used with metadata associated is available at 
\faIcon{github}~\url{https://github.com/paolonapoletano/bci-resources}.
}
\label{fig:data}
\end{figure*}

\subsection{Data Scarcity}

The rise of LBM models as dominant approaches has underscored the growing need for large, high-quality datasets.  
To assess the feasibility of applying LBM models to EEG-BCI data, we collected metadata on datasets published from 2004 to the present, covering six EEG-BCI tasks: Brain-to-Image (B2I), Brain-to-Text (B2T), Emotion Recognition (ER), Motor Imagery (MI), Neurodegenerative Disease (ND), and Seizure Prediction (SP). Figure~\ref{fig:data} presents key dataset statistics, including total recorded time per task, and technical details of the sampling process, such as the number of EEG channels and sampling frequency.  
A detailed version of the metadata used in this analysis is available here~\footnote{
\faIcon{github}~\url{https://github.com/paolonapoletano/bci-resources}
}.  

As it can be seen in Figure~\ref{fig:data_time}, despite the large number of datasets currently available, the total recorded time is limited, particularly for certain tasks such as Brain-to-Image and Brain-to-Text. Additionally, Figures~\ref{fig:data_samplingfreq} and \ref{fig:data_channels} reveal trends in specific tasks regarding sampling frequency, resolution, and the type of technology used for recording. Moreover, datasets exhibit stark heterogeneity, making it difficult to use multiple datasets simultaneously for a specific task.
Another important aspect is the variation in the number of subjects recorded across different tasks. As shown in Figure~\ref{fig:data_time}, the total number of subjects per task varies significantly, revealing a substantial disparity. This imbalance limits the ability to study intra- and inter-subject dependencies across recordings.

Beyond the EEG data analysis just presented, we also seek to quantify the available data volume and compare it to that of the Natural Language Processing field, as current LBM models (see Section \ref{sec:ssl}) depend on such data for pre-training. To this end, we approximate the number of available tokens for a given task as in the following.

Let \(T_{i}\) be the total recording time (in seconds) for a task $i$, and \(C_i\) the corresponding average number of EEG channels; let \(L\) be the token duration in seconds (e.g., \(L=1\) sec, as in \citep{jianglarge}). The approximate number of tokens for a task can be thus derived as $\mathcal{N}_{tokens}(i) = (T_i \times C_i)/L$.

Table~\ref{tab:tokens} provides the overall estimated values, accounting for all publicly available datasets for each task in terms of total recorded time ($T_i$), number of EEG channels ($C_i$), and number of tokens ($\mathcal{N}_{tokens}(i)$).

These results highlight that LBMs are trained on datasets containing at most millions of tokens ($10^6$) if we consider all data across tasks. In contrast, a LLM model like LLaMA 3.1 405B has been trained on 15 trillion ($10^{12}$) tokens~\footnote{\url{https://ai.meta.com/blog/meta-llama-3-1/}}. This clearly illustrates how limited is the volume of available EEG data if compared to the Natural Language Processing field.

\begin{table}[tb]
    \centering
    \caption{Estimation of the number of tokens available for LBM training, extracted from datasets per task. The total time is calculated as the sum of recorded hours for each task, regardless of differences between datasets.  For comparison, a LLM model like LLaMA 3.1 405B has been trained on 15 trillion ($10^{12}$) tokens.}
    \label{tab:tokens}
    \adjustbox{width=0.6\linewidth}{
    \begin{tabular}{lccc}
        \toprule
        Task   & Total Time     & \# channels  & \# of tokens \\ 
           & $T$ (h)     &  $C$ (avg)  & $\mathcal{N}$ \\ \midrule
        Brain-to-Image              &$\sim$ 375     & 61    & 81M    \\
        Brain-to-Text               &$\sim$ 494     & 60    & 108M   \\
        Emotion Recognition         &$\sim$ 2560    & 44    & 406M   \\
        Motor Imagery               &$\sim$ 984     & 50    & 179M   \\
        Neurodegenerative Diseases  &$\sim$ 1273     & 35    & 161M    \\
        Seizure Prediction          &$\sim$ 2969    & 18    & 193M   \\
        \bottomrule
    \end{tabular}
    }
    
\end{table}

\subsection{Lack of Interventional Data}

Beyond the mere need to scale the amount of EEG data, we argue that if the ultimate goal is to develop EEG-based BCIs for real-world deployment, datasets must encompass the most diverse set of conditions possible. From a causal reasoning perspective, this translates into collecting data under targeted interventions on the model depicted in Figure \ref{fig:map}. Achieving this requires rethinking EEG dataset collection in a more controlled experimental fashion, where a specific variable is manipulated while all others remain fixed.

Notably, such data would not only provide a widely diversified training set for AI-based BCI models but also enable proper evaluation and benchmarking of different approaches in terms of their interventional robustness. When only observational data is available, we can evaluate models solely based on their predictions within the observed data distribution, but we cannot assess their behavior under interventions. How effective is my BCI in translating Motor Imagery into actual movement when the subject changes? What if the subject is at home rather than in a university lab? Will the model still work if I switch to a different EEG device? These are fundamental yet unanswered questions if only observational data is available. 

A similar concern has been raised in cognitive neuroscience regarding the assessment of models describing the neuronal underpinnings of cognitive functions \citep{weichwald2021causality}. In line with this, we argue that appropriate benchmarking procedures should also be employed to evaluate EEG-based BCI models. In this context, the availability of interventional data would enable the adoption of \emph{leave-one-intervention-out} testing paradigms to assess a model’s invariance to specific distribution shifts.

To some extent, this approach has already been applied to a subset of easily collectible shifts, typically subject and session shifts (cfr. Section \ref{sec:context}), but overall, this remains a largely unexplored territory.

\section{Future Trends and Conclusive Remarks}
\label{sec:trends}

Our exploration thus far has forged a conceptual map  for charting the confluence of EEG-based BCI and recent advances in AI/ML, while highlighting the grand challenges that lie ahead. In turn, these very challenges expose the technological and methodological gaps we will now examine and discuss in the remainder of this conclusive Section.  We argue
that addressing these gaps is essential for advancing EEG-
based BCIs toward practical, real-world applications. Indeed, the endeavor to overcome these obstacles will likely define the forthcoming research directions. 

\subsection{Bridging the Technological Gap}
Despite significant advancements in EEG-based BCIs, current technology still faces limitations that hinder the development of portable applications. One major challenge concerns the reliability of dry electrodes compared to wet electrodes. While dry electrode systems, such as OpenBCI Ultracortex~\footnote{\url{https://openbci.com/}} and EMOTIV Epoc X~\footnote{\url{https://www.emotiv.com/}}, offer greater ease of use and reduced setup time, they suffer from lower signal quality and greater susceptibility to noise compared to wet electrode systems like g.tec g.Nautilus~\footnote{\url{https://www.gtec.at/}} or Brain Products LiveAmp~\footnote{\url{https://www.brainproducts.com/}}. As a middle ground, salt-based electrodes, such as those used in g.tec g.SAHARA, provide better signal quality than dry electrodes without requiring conductive gels. However, they still struggle with signal stability over time due to evaporation and the need for periodic rehydration.

Another key challenge is the computational demand of modern AI models. Most deep learning techniques require substantial processing power, making them impractical for mobile and wearable BCI systems. The integration of Edge AI, which optimizes models for on-device processing, is crucial for enabling real-time, low-latency inference while maintaining power efficiency. Recent efforts to implement AI models on embedded hardware, such as NVIDIA Jetson platforms~\footnote{\url{https://www.nvidia.com/}} and low-power AI accelerators~\footnote{\url{https://stm32ai.st.com/}}, represent promising steps toward overcoming this limitation.

Beyond hardware and computational constraints, there is also a  gap in the standardization and reproducibility of EEG analysis. The lack of benchmarking tools and widely accepted datasets makes comparing AI approaches and validating experimental results across studies difficult. Initiatives like the Mother of All BCI Benchmarks (MOABB)~\citep{Aristimunha_Mother_of_all_2023} are a step toward establishing standardized evaluation protocols, promoting reproducibility, and facilitating the adoption of AI methods in EEG-based BCI research. However, further progress is needed to enhance transparency, encourage data sharing, and improve cross-study comparisons.

\subsection{Bridging the methodological gap}

\subsubsection{Modeling the Temporal Dynamics of EEGs}

One of the fundamental challenges of EEG-based BCI is the accurate modeling of the temporal dynamics inherent in neural activity. EEG signals exhibit highly non-stationary and multiscale temporal structures, including rapid and transient responses (e.g., ERPs) and sustained oscillatory patterns over different frequency bands~\citep{roy2019towards}. However, most deep learning approaches, including CNNs and RNNs, struggle to fully capture these intricate dynamics because of their inherent limitations in handling long-range temporal dependencies and fine-grained temporal structures.  

In addition, transformer-based models, including LLMs adapted to time series data, have demonstrated strong feature extraction capabilities but remain limited by their dependence on large-scale training data and computationally expensive self-attention mechanisms, which may not efficiently capture the sparse, event-driven nature of EEG~\citep{ren2021reconstructing} signals.

Some have proposed biologically inspired neural networks that learn without supervision—such as Adaptive Resonance Theory (ART) \citep{Grossberg2018ART}—aiming to achieve autonomous adaptive intelligence. These methods seek to explain how the brain self-organizes representations while adapting to new information in a dynamic environment, a fundamental capability for addressing the generalization challenges of current BCI approaches.

In a different vein, spiking neural networks (SNNs) represent a significant alternative in the artificial intelligence landscape, particularly for their application to BCI systems using EEG signals~\citep{wang2020supervised}. Unlike traditional artificial neural networks, which rely on continuous activation functions, SNNs process information using discrete spikes that more closely emulate biological neurons' behavior~\citep{dora2021spiking}. This spike-based paradigm increases the biological plausibility of artificial networks. It enables more efficient integration with neuromorphic hardware platforms with significantly lower power consumption than traditional hardware platforms such as GPUs and CPUs~\citep{kumar2022decoding}. Low power consumption is essential for portable devices, such as BCI wearables or mobile diagnostic tools, which often run on battery power. The increased energy range allows these systems to operate for extended periods without recharging, making them ideal for mobile real-time applications. 

The temporal encoding of SNNs offers a natural advantage in the analysis of EEG signals, as they are well suited for processing time-varying data. By encoding information in the timing of spikes, SNNs can capture the intricate temporal dynamics of EEG signals, including event-related potentials (ERPs) and oscillatory activities over different frequency bands~\citep{roy2019towards}. In addition, SNNs offer greater efficiency in handling sparse data, a common feature of EEG signals after preprocessing steps such as artifact removal or dimensionality reduction~\citep{roy2019towards}. Recent studies have demonstrated the potential of SNNs in tasks such as motor imagery classification~\citep{antelis2020spiking}, cognitive state detection~\citep{li2024spiking}, and epilepsy monitoring~\citep{tian2021new}, showing competitive performance with conventional deep learning methods while requiring significantly fewer computational resources.

Despite the promise, the adoption of SNNs and neuromorphic processing in EEG-based BCIs is not without challenges. Training SNNs remains a complex task due to the lack of widely adopted learning algorithms, particularly for supervised tasks. Although Spike-Timing Dependent Plasticity (STDP) and surrogate gradient methods have shown promise, further research is needed to develop scalable and robust training techniques~\citep{tavanaei2019deep}.

\subsubsection{Next-Gen Large Brain Models}

As a matter of fact, when inspecting the most recent proposals for EEG-based BCIs, it becomes clear that most of the attention is devoted to leveraging the successes of Large Language Models for modeling brainwave data. However, slavishly retracing the steps that have made modern language modeling so successful may hinder the advancement of LBMs.

We argue that, when comparing the two fields, a stark difference emerges which relates to the structure of the data being modeled. For instance, while discrete tokenization is well-suited for text processing, applying it to brainwave signals— which have a continuous-time nature— may be less appropriate. In this regard, it could be interesting to explore recent approaches aimed at "reasoning" in continuous time \citep{hao2024training}.

Another important consideration arises from the discussion presented in this paper. Causal reasoning may represent one of the most promising strategies for addressing the longstanding challenge of BCI generalization. In this respect, it is worth noting that the most recent language processing literature is beginning to develop methods that integrate LLMs with causal reasoning. Indeed, empirical evidence suggests that large-scale pretraining alone is insufficient for enabling models to encode causal knowledge or reasoning; at best, they can recite it if explicitly present in the training data—hence the characterization of LLMs as "causal parrots" \citep{zevcevic2023causal}. Consequently, some efforts are being devoted to developing causality-aware LLMs (see \citep{wu2024causality} for a comprehensive review). We envision EEG-based BCIs grounded in LBMs making a similar transformative leap.

\subsection{Ethical and Societal Implications of AI-Based BCI Systems}
Developing and deploying AI-based EEG-BCI systems raises several important ethical and social issues~\citep{ng2025exploring,burwell2017ethical}. 

A central concern is mental privacy, since brain-computer interfaces can provide access to neural activity that may reflect private thoughts, intentions, or emotional states~\citep{schroder2025cyber}. Unlike other forms of biometric data, brain signals are highly sensitive and are not yet adequately protected by an existing legal framework~\citep{ligthart2023minding}. Cybersecurity vulnerabilities present a significant challenge. Modern BCI systems that rely on wireless or network connectivity are susceptible to unauthorized access, data manipulation, and malicious interference~\citep{kritika2024comprehensive}. Addressing these risks requires stringent security measures, including strong encryption, authenticated communication protocols, and secure mechanisms for software updates~\citep{schroder2025cyber}.

Another critical concern is user autonomy. In particular, closed-loop or assistive AI-based BCI systems may influence—or even override—the user's intentions, emotional states, and decision-making processes~\citep{kellmeyer2016effects}. This issue is further complicated by the lack of explainability in most AI models, making it difficult for users and clinicians to fully understand or challenge the system’s decisions, especially in clinical or high-risk environments.  These factors raise ethical concerns about the safety and effectiveness of such technologies, particularly when used for enhancement in healthy individuals, where the potential risks may outweigh the benefits~\citep{ng2025exploring}.

\newpage

\section*{Disclosure statement}
No potential conflict of interest was reported by the authors.

\section*{OrcID}

Thomas Barbera \url{https://orcid.org/0009-0001-5684-1867}\\
Jacopo Burger \url{https://orcid.org/0009-0007-3048-2038}\\
Alessandro D’Amelio \url{https://orcid.org/0000-0002-8210-4457}\\
Simone Zini \url{https://orcid.org/0000-0002-8505-1581}\\
Simone Bianco \url{https://orcid.org/0000-0002-7070-1545}\\
Raffaella Lanzarotti \url{https://orcid.org/0000-0002-8534-4413}\\
Paolo Napoletano \url{https://orcid.org/0000-0001-9112-0574}\\
Giuseppe Boccignone \url{https://orcid.org/0000-0002-5572-0924}\\
Jose Luis Contreras-Vidal \url{https://orcid.org/0000-0002-6499-1208}

\section*{Funding}
This work was partially funded by the National Plan for NRRP Complementary Investments (PNC, established with the decree-law 6 May 2021, n. 59, converted by law n. 101 of 2021) in the call for the funding of research initiatives for technologies and innovative trajectories in the health and care sectors (Directorial Decree n. 931 of 06-06-2022) - project n. PNC0000003 - AdvaNced Technologies for Human-centrEd Medicine (project acronym: ANTHEM). This work reflects only the authors’ views and opinions, neither the Ministry for University and Research nor the European Commission can be considered responsible for them. The publication was carried out with co-funding from the Ministry of University and Research within the framework of the PNC at the Department of Computer Science ``Giovanni degli Antoni'', University of Milan – Via Celoria 18, Milan, and the U.S. National Science Foundation (NSF) IUCRC BRAIN award n. 2137255 at the University of Houston.

\bibliographystyle{apacite}
\bibliography{bibliography}

\end{document}